\documentstyle[12pt]{article}
\begin{document}
\begin{flushright}
{\bf DO-TH 94/15 \\
November 1994}
\end{flushright}
\vskip .5in
\begin{center}
{\Large Baryogenesis from a Lepton Asymmetric Universe}\\
\vskip .75in
{\large Marion Flanz, Emmanuel A. Paschos and Utpal Sarkar\footnote{
{\it On sabbatical leave from} Theory Group, Physical Research Laboratory,
Ahmedabad - 380009, India.}}
\vskip .4in
\baselineskip 20pt
{\large Institut f\"{u}r Physik},\\ {\large Universit\"{a}t Dortmund,}\\
{\large D-44221 Dortmund, Germany.}
\vskip 1in
\end{center}
\begin{abstract}
\baselineskip 20pt

We demonstrate that $CP-$violation in the Majorana mass matrices of
the heavy neutrinos can generate a $CP-$asymmetric universe. The
subsequent decay of the Majorana particles
generates a lepton number asymmetry. During the electroweak phase transition
the lepton asymmetry is converted into a baryon asymmetry,
which survives down to this time.

\end{abstract}

\newpage
\baselineskip 20pt

The cosmological  baryon
excess can be generated from the initial condition $B=0$ through
the baryon number violating decays of very heavy particles  \cite{kolb}.
This requires violation of both $C$ and $CP$ and, in addition,
these processes
must happen during an epoch which brings the universe out of equilibrium.
This mechanism has two major drawbacks. In most grand unified theories,
baryon changing decays conserve $(B-L)$ and anomalous electroweak
interaction induced by sphalerons can wash out the asymmetry \cite{anom}.
Second, the asymmetry generated in most models is much smaller than the
ratio $$
\frac{n_B}{n_\gamma} = (4 - 7) \times 10^{-10} $$ required by
observations and big-bang cosmology. A second approach relies
on the violation of baryon number through the global $(B-L)-$anomaly which
together with $C-$ and $CP-$violation generates a baryon excess
at the electroweak phase transition, provided it is first order
\cite{brahm}. The search for a consistent extension of the
standard model is under active investigation.

At present, a very attractive scenario consists of generating a lepton
number asymmetry through lepton number violating decays of heavy neutrinos.
The baryon asymmetry is then generated from the lepton asymmetry
during the electroweak phase transition, mentioned above.
Some time ago, Fukugita and Yanagida proposed this mechanism
[4-7]
where the lepton number violation was introduced explicitly
through a Majorana mass term of the right-handed neutrinos.
In all these models
the $CP-$violation is introduced through the interference of
tree-level and one-loop diagrams in the decays of the heavy
neutrinos.

Attempts were made \cite{xx1,xx2,xx3} to generate the baryon asymmetry
in models, where the $CP-$violation comes through the mixing of scalars.
The possibility of mixing scalar particles in order to obtain
large $CP-$violation
was discussed some time back \cite{xx1}. A similar idea
was recently exploited in the context of GUTs \cite{xx2,xx3}, where
the oscillation of scalar particles was proposed.
In these scenarios, the GUTs were extended to include new scalars which
allow such interactions and introduce $\epsilon$-type
$CP-$violating effects in the mass matrix of the scalars.
In the above models $CP-$ and baryon-violation occur in the same
diagrams.
In a recent article \cite{pas}, $\epsilon$-type effects
have been considered in a different way. Through the scattering
of heavy bosons it has been shown that it is
possible to create first a $CP-$asymmetric universe (to be precise
$CP-$asymmetric density of scattering states) before the
heavy particles decay.
When they begin to decay they generate the baryon
asymmetry.
The main difference of this mechanism from the previous ones lies
in the origin of $CP-$violation. In the latter mechanism \cite{pas} the
$CP-$violation arises from the phases in the couplings of the
matter--anti-matter oscillations in which a quantum number
changes by two units, in contrast to references \cite{xx1,xx2,xx3} where
the $CP-$violation occurs in decays of
the heavy particles and the quantum numbers change
by one unit. If this oscillation violates baryon
number by two units, and at the same time there is $CP-$violation,
then the oscillation will generate a $CP-$asymmetric and
also baryon asymmetric universe.

In the case of heavy Majorana neutrinos, there exists one loop
self-energy diagrams introducing a correction to the
Majorana mass of the particles. These
diagrams will not conserve $CP$ and will produce neutrino--anti-neu\-tri\-no
oscillations violating lepton number by two units, which can
contribute to the lepton number asymmetry of the universe. In
the case of three generation models, the $CP-$violating phases
which enter in the self energy diagrams are usually
not the ones appearing in the
interference terms of the decays of the heavy neutrinos \cite{cpnu}.
Thus this contribution of $C-$, $CP-$ and $(B-L)-$violation to
leptogenesis is distinct from the ones considered in the literature.
In addition, the contributions are larger than the earlier ones for
several choices of parameters.

We consider a theory with the same particle content as in
ref \cite{fy}, that is, the particle content of the standard
model plus three Majorana neutrinos $ N_i $ with $i = 1,2,3 $,
one for each generation. The new Majorana neutrinos are singlets under
the standard model, i.e.
$SU(3)_c \times SU(2)_L \times U(1)_Y$.
The new interactions of these neutrinos with leptons and the Majorana
mass term are given by,
\begin{equation}
{\cal L} = M_{ij} {\bar N}_i^c N_j
+ h_{\alpha i} {\bar N}_i l_{\alpha} \phi^\dagger + {\rm h.c.}
\end{equation}
where $\phi$ is the standard model higgs doublet, which breaks the
electroweak symmetry and gives mass to fermions.

The lepton number and $(B-L)$ are
violated through the Majorana mass of the
right-handed neutrinos $N_i$'s. In all other interactions
lepton number is conserved. We shall work in the density matrix
formalism \cite{ellis} to estimate the contribution  of
neutrino--anti-neutrino
oscillations to the lepton number generation of the universe. The
neutrino state $|N>$ and the anti-neutrino state $|N^c>$ can be
distinguished by their decay properties. The state $|N>$ decays into a
light lepton $l_i$ and a higgs $\phi^\dagger$, whereas the state $|N^c>$
decays into a light anti-lepton $l_i^c$ and a higgs $\phi$.
The only interaction which relates the two states and violates
lepton number is the Majorana mass term, which also causes mixing
of the two states. We shall discuss the question of $CP-$violation
in such a mixing later on. We shall assume $CPT-$invariance,
but violation of $CP$. For the sake of simplicity, we consider two
generations of Majorana neutrinos where the indices $i$ and $j$
take the values 1 and 2. We define a basis of states as
$(N_1^c \:\:\:\: N_2^c \:\:\:\: N_1 \:\:\:\: N_2 )$ and the
Hamiltonian takes the form,
$$
\pmatrix{0 & 0 & H_{11} & H_{12} \cr 0 & 0 & H_{12} & H_{22} \cr
H_{11} & H_{21} & 0 & 0 \cr H_{21} & H_{22} & 0 & 0 }. $$
Without loss of generality, we work in the basis where
the Majorana mass matrix is diagonal
and real. Then the terms $H_{11}$ and $H_{22}$ are the mass terms in
equation (1). In this basis, the element $H_{12}$ is
generated through the diagrams of figure 1. Similar diagrams
produce $H_{21} = M_{21} - \frac{i}{2} \Gamma_{21}$ with
$M_{21}=M_{12}^* \:,\:\:\: \Gamma_{21}=\Gamma_{12}^*$.

We define the $4 \times 4$ density matrix in the basis of states
defined above. It satisfies the equation,
$$ \frac{d \rho}{dt} = -i (H \rho - \rho H^\dagger). $$
The solutions are obtained by diagonalizing $H$. The expectation
value of any observable,
$$ {\cal O}_f = <f| \pmatrix{|N_1^c><N_1^c| & |N_1^c><N_2^c| &
... & |N_1^c><N_2| \cr |N_2^c><N_1^c| &
... & ... & ... \cr ... &
... & ... & ... \cr |N_2><N_1^c| &
... & ... & |N_2><N_2|  }
|f>,$$ is given by,
\begin{equation}
< {\cal O}_f > = \displaystyle \frac{{\rm Tr} ( {\cal O}_f \rho )}{
{\rm Tr} \rho} \, ,
\end{equation}
where $|f>$ is the final states of leptons or anti-leptons produced
through the decays of these
density of state $|N> \to l \phi^\dagger$ and $|N^c> \to l^c \phi$.
The observables corresponding to leptons and antileptons in the final
states are, respectively,
$$
{\cal O}_{l_\alpha} \propto \pmatrix{0 & 0 & 0 & 0 \cr
0 & 0 & 0 & 0 \cr 0 & 0 & h_{\alpha 1} h^*_{\alpha 1} &
h_{\alpha 1} h^*_{\alpha 2} \cr  0 & 0 & h_{\alpha 2} h^*_{\alpha 1} &
h_{\alpha 2} h^*_{\alpha 2} }  \quad and \quad
{\cal O}_{\bar{l}_\alpha} \propto \pmatrix{
h_{\alpha 1} h^*_{\alpha 1} & h_{\alpha 1} h^*_{\alpha 2} & 0 & 0 \cr
h_{\alpha 2} h^*_{\alpha 1} & h_{\alpha 2} h^*_{\alpha 2} & 0 & 0  \cr
0 & 0 & 0 & 0 \cr 0 & 0 & 0 & 0  }. $$
Hence a lepton asymmetry is generated through the $CP$-violating
phase and is given by,
\begin{equation}
\delta = \frac{\Gamma_l - \Gamma_{\bar{l}}}{\Gamma_l
+ \Gamma_{\bar{l}}}.   \label{eq3}
\end{equation}

The above asymmetry $\delta$ calculated later on
is different than the asymmetry generated from the decay of heavy neutrinos
\cite{xx3}, because the elements of the density matrix already contain
$CP$-violation. In the terminology of $K-$meson decays it
originates in the mass-matrix, that is
the density matrix described above (similar to that of
ref.\cite{pas})\footnote{We thank Dr. H. So for pointing out that the
specific mechanism in ref.\cite{pas} is suppressed by the general property of
GUTs, according to which $(B-L)$-changing operators are suppressed
relative to the $(B+L)$-changing operator. In ref. \cite{pas} the
suppression is implied by the conservation of weak isospin.}.

In the present case, the physical states are pure states and can also
be described in the wave function formalism. For the mass matrix
under consideration, we give the mass eigenstate for one
of the eigenvalues as,
$$
\psi_1 \approx |N_1 + N_1^c > + b | N_2 > + d | N_2^c > $$
The decay widths for the decays of $\psi_1$ into leptons and antileptons
is given by,
\begin{eqnarray}
\Gamma_l &\propto& | h_{\alpha 1} + b \:\: h_{\alpha 2} |^2 \nonumber \\
\Gamma_{\bar{l}} &\propto& | h^*_{\alpha 1} + d \:\: h^*_{\alpha 2}
|^2 ,  \nonumber
\end{eqnarray}
with,
\begin{eqnarray}
b &\approx& \frac{H_{12} H_{11} + H_{21} H_{22} }{
H^2_{11} - H^2_{22}}  \nonumber \\
{\rm and} {\hskip .3in} d &\approx& \frac{H_{21} H_{11} + H_{12}
H_{22} }{H^2_{11} - H^2_{22}} \nonumber
\end{eqnarray}
derived from the Hamiltonian perturbatively.
In a similar way we can now write down the decay rates for the states
$\psi_2, \psi_3 $ and $\psi_4$. The quantities $b$ and $d$
are determined form the loop diagram in figure 1. The final asymmetry
obtained either in the density matrix
or the wave function formalism. The oscillation of a $j=2$
neutrino to an $i=1$ neutrino and its decay to the $\alpha$th
light lepton or antilepton produces the asymmetry,
\begin{equation}
\delta = 4 \sqrt{2} \pi \left( \frac{M_{i}}{M_{j}} \right)
\frac{ {\rm Re}\:\:(h_{\alpha j}
h^*_{\alpha i})\:\:\:{\rm Im}(h_{\alpha j} h^*_{\alpha i}) }{
\left| h_{\alpha i} \right|^2 + \left| h_{\alpha j} \right|^2 }
\end{equation}
We shall include this lepton asymmetry in the evolution
of the lepton number determined by
the Boltzmann equation.

The number density of the left-handed leptons $n_l$ will
evolve in time following the equation,
\begin{equation}
\frac{d n_l}{d t} + 3 H n_l = \left[
\Delta_l n_j \Gamma_i + \frac{1}{2} (1 + \epsilon^\prime) n_i \Gamma_i
- {\cal D}_{i}  \right]  \label{eq4}
\end{equation}
The second term on the left side of the equation comes
from the expansion of the universe, where
$H$ is Hubble's constant. We now discuss the origin of the
various terms on the right of this expression.

The first term contains the number density $n_j$ of initial
states $|N_j>$, which is converted to the
final state $|N_i^c>$,
which subsequently decays into light leptons with a decay width
$\Gamma_i = \frac{h_{ii}^2}{16 \pi^2} M_i$. Here $M_i$
is the Majorana mass eigenstate of the $i$-th neutrino.
The factor $\Delta_l$ contains the $CP-$conserving and
$CP-$violating terms of the $|N>-|N^c>$ oscillations.
The second term in the expression $(\ref{eq4})$
arises from the $CP-$violation
in the decays of the heavy leptons. The interference of the tree
level diagram and the one loop penguin--type diagrams in the
decays of the neutrinos $N_i$ contains the $CP-$violating phase
and the absorptive part of the integral. This contribution is the
usual one which was studied in the literature in detail \cite{fy,luty,us}.
The measure of the $CP-$violation is now given by $\epsilon^\prime$.
In the absence of the new contribution, {\it i.e.,} of the
term proportional to $\Delta_l$, the asymptotic value of the
lepton number asymmetry is
proportional to $\epsilon^\prime/g_*$.

The evolution of the light anti-leptons will be given by
similar equations, with $\Delta_l \to \Delta_{\bar{l}}$,
$\epsilon^\prime \to - \epsilon^\prime$ and $l \to \bar{l}$.
This way we arrive at a Boltzmann equation for the lepton difference
$n_L s = n_l - n_{\bar{l}}$, where $s = g_* n_\gamma$ is the
entropy density, $g_*$ the number of active degrees of freedom
and $n_\gamma$ is the number of photons. In the difference occurs
a term proportional to $\delta = \Delta_l - \Delta_{\bar{l}}$
computed in equation (\ref{eq4}).

There are also various processes which deplete the number
densities of the left-handed leptons. They come from the reactions
$l + \phi^+ \to l^c + \phi$. We can denote by $D_l$ the depletion of
lepton and $D_{\bar{l}}$ the depletion of the antileptons. The
difference which occurs in the asymmetry equation is,
\begin{equation}
{\cal D} = D_l - D_{\bar{l}} = (2 \delta + \epsilon^\prime)
n^{eq} (\Gamma_i + \tilde{\sigma_v}) +
n^{eq} (\Gamma_i + \tilde{\sigma_v}) \frac{n_L}{n_\gamma}
\end{equation}
The term with $\Gamma_i$ comes from the $|N>$ or $|N^c>$ neutrinos in
the intermediate state and $\tilde{\sigma_v}$ is produced from the remaining
intermediate states. Since this contribution is, in general, smaller
than $\Gamma$ it will be neglected. Direct and indirect $CP-$violation
in the intermediate states produce the first term. The difference
$|{\cal A}(l \phi^c + \to l^c \phi) |^2 -
|{\cal A}(l^c \phi + \to l \phi^c) |^2$
will also contribute to the depletion through the term proportional to
$\frac{\mu_l - \mu_{\bar{l}} }{T} = \frac{n_l - n_{\bar{l}} }{n_\gamma}$.
When $\tilde{\sigma_v}$ is neglected,
the last term in this expression is due to the
recombination of the left-handed leptons and the higgs to form the
heavy neutrinos. All these processes are rapid enough to occur in
thermal equillibrium and hence the number density entering here
is the equilibrium number density $n^{eq}$ of $N$.

We now introduce $Y_i = n_i/s$, $Y^{eq} = n^{eq}/s$,
the dimensionless variable, $x = M_1/T$ and make use of the relations,
$t = 1/2 H = x^2 / 2 H(x =1)$ with $H(x = 1) = 1.73 \sqrt{g_*}
\frac{M_{1}^2}{M_{Pl}}$ (from now on we assume that
$M_{1}<M_{2}<M_{3}$ and the lepton number asymmetry is generated just
before $T \sim M_{1}$). We also define the parameter,
$$ K = \displaystyle \frac{\Gamma_i(x=1)}{H(x=1)},$$
which is the measure of the out-of-equilibrium condition.
For the usual extensions of the standard model, one can
take $g_* = 400$. With these redefinitions we obtain a
simplified equation for the evolution of the lepton number
asymmetry of the universe,
\begin{equation}
\displaystyle \frac{{\rm d} n_L}{{\rm d }x} = K x^2 \left[
(2 \delta  + \epsilon^\prime) (Y_i - Y^{eq})
- Y^{eq} g_* n_L \right]. \label{nl}
\end{equation}
The $Y_i$ satisfy the Boltzmann equations,
\begin{equation}
\displaystyle \frac{{\rm d} Y_i}{{\rm d }x} = - K x^2 (Y_i - Y^{eq}).
\end{equation}

To solve these equation we assume that although the universe is
out-of-equilibrium, it stays close to it, so that $\frac{{\rm d}
(Y_i - Y^{eq})}{{\rm d} x} = 0$. Then it is easier to solve
the equations and obtain an estimate of the lepton number generated.
For very large time, the solution of a differential equation
has an asymptotic value, which is almost independent of the constant
$K$ for $1 \leq K \leq 50$, and is given approximately by,
\begin{equation}
n_L = \frac{1}{g_*} (2 \delta  + \epsilon^\prime).
\end{equation}
If $\delta = 0$, we recover the expression for the
scenarios \cite{fy,luty}, where the heavy neutrinos can only decay.
We compare next the two $CP-$violating contributions in these
scenarios.

The $CP-$violation in the decays of heavy neutrinos has
been discussed in the literature and for the model under consideration
it is given by \cite{fy,luty,us},
\begin{equation}
\epsilon^\prime = \frac{1}{2 \pi} \frac{{\rm Im} (h_{\alpha 1}
h_{\alpha j} h^*_{\beta 1} h^*_{\beta j})}{h_{\alpha 1}
h^*_{\alpha 1}} \frac{M_1}{M_j} .
\end{equation}
{}From this expression it is clear that the present contribution
$\delta$  is different from the contribution considered earlier,
given be $\epsilon^\prime$. There exists various choices of parameters
for which $\delta \gg \epsilon^\prime$.
In all these cases, the matter--anti-matter
oscillations will first create a $CP-$asymmetric universe with
unequal densities of Majorana neutrinos and antineutrinos,
whose decays generate a lepton asymmetric universe.

In the presence of sphalerons, this lepton number asymmetry
will be converted to the baryon number asymmetry.
After the electroweak phase transition, the baryon number of the universe
will be given by,
$$ n_B = - \left( \displaystyle \frac{8
N_g + 4 N_H}{22 N_g + 13 N_H} \right) n_{(B-L)}
\sim \frac{1}{3} n_L \sim \frac{2 \delta }{3 g_*}, $$
where $ N_H $ is the number of Higgs doublets and $ N_g $ the number of
generations.
For $h_{13} \sim h_{11}$ and $g_* \sim 400$, the required amount of
baryon asymmetry can be generated with a $CP-$violating phase
$\delta \sim 10^{-5} - 10^{-7}$, which is a very natural choice.

We have shown that $CP-$violation in the mass matrix of heavy Majorana
neutrinos produces a $CP$ and lepton asymmetric universe. The same mass
terms break the $(B-L)$ symmetry and produce a net excess of this quantum
number which survives all the way down to our epoch. The asymmetries created
on the mass matrix are relatively large and are transformed into an excess
of Baryons, when the electroweak phase transition takes place.

\vskip .3in
{\bf Acknowledgement} One of us (US) would like to acknowledge
a fellowship from the Alexander von Humboldt Foundation and
hospitality from the Institut f\"{u}r Physik, Univ Dortmund
during his research stay in Germany. We would like to thank
Drs. A. Pilaftsis and H. So for several discussions. The
financial support of BMFT is greatfully acknowledged (05-6DO93P).

\newpage
\begin{figure}[htb]
\mbox{}
\vskip 6in\relax\noindent\hskip .5in\relax
\includegraphics{dof1.ps}
\caption{ One loop diagrams contributing to the
$CP$ violating $|N_1> \to |N_2^c>$ transition.}
\end{figure}


\newpage
\begin{thebibliography}{99}
\baselineskip 16pt
\bibitem{kolb} A.D.Sakharov, Pis'ma Zh.Eksp. Teor. Fiz. {\bf 5}
  (1967) 32; E.W.Kolb and M.S.Turner, {\it The Early Universe}
  (Addison-Wesley, Reading, MA, 1989).

\bibitem{anom} G. 't Hooft, Phys. Rev. Lett. {\bf 37} (1976) 8;
  V.A. Kuzmin, V.A Rubakov and M.E. Shaposhnikov,  Phys. Lett.
  {\bf B 155} (1985) 36.

\bibitem{brahm} D.Brahm, in {\it Proc. XXVI Int. Conf. H.E.
  Physics}, Dallas, August 1992.

\bibitem{fy} M. Fukugita and T. Yanagida, Phys. Lett. {\bf B 174}
  (1986) 45.

\bibitem{luty} M.A.Luty, Phys.Rev. {\bf D 45} (1992) 445; P.Langacker,
  R.D.Peccei and T.Yanagida, Mod.Phys.Lett. {\bf A1} (1986) 45;
  R.D. Peccei, in {\it Proc. XXVI Int. Conf. H.E.
  Physics}, Dallas, August 1992;
  H.Murayama, H.Suzuki, T.Yanagida and J.Yokoyama,
  Phys. Rev. Lett. {\bf 70} (1993) 1912.

\bibitem{us} A.Acker, H.Kikuchi, E.Ma and U.Sarkar, Phys. Rev.
  {\bf D48} (1993) 5006.

\bibitem{pat} P.J. O'Donnell and U. Sarkar, Phys. Rev. {\bf D 49}
  (1994) 2118.

\bibitem{xx1} A. Yu. Ignatev, V.A. Kuzmin and M.E. Shaposnikov, JETP
  Lett. {\bf 30} (1979) 688.

\bibitem{xx2} F.J. Botella and J. Roldan, Phys. Rev.
  {\bf D 44} (1991) 966.

\bibitem{xx3} J. Liu and G. Segre, Phys. Rev. {\bf D 48} (1993) 4609.

\bibitem{pas} M. Flanz, E.A. Paschos, Y.L. Wu, Phys. Lett.
  {\bf B 315} (1993) 379.

\bibitem{cpnu} B. Kayser, in {\it CP violation}, ed. C. Jarlskog
  (World Scientific, 1989) p. 334; J. Schechter and J.W.F. Valle,
  Phys. Rev. {\bf D 22} (1980) 2227; J.F. Nieves and P.B. Pal,
  Phys. Rev. {\bf D 36} (1987) 315.

\bibitem{ellis} J.E.Ellis, J.S.Hagelin, D.V.Nanopoulos and M.Srednicki,
  Nucl.Phys. {\bf B241} (1984) 381; J.E.Ellis, N.E.Mavromatos and
  D.V.Nanopoulos, Phys.Lett. {\bf B293} (1992) 142; CERN preprint
  CERN-TH.6755/92.

\end{thebibliography}
\end{document}